\documentclass[preprint]{aastex}
\usepackage{epsf}

\newcommand{\thir}{\mbox{$\Delta I_{\mbox{\tiny 30}}$}}
\newcommand{\yc}{\mbox{$y_{\mbox{\tiny 0}}$}}
\newcommand{\dec}{\mbox{$\Delta I_{\nu\mbox{\tiny 0}}$}}
\newcommand{\vp}{\mbox{$v_{\mbox{\scriptsize {\it p}}}$}}

\begin{document}

\title{The Sunyaev-Zel'dovich Effect Spectrum of Abell 2163}
\author{
\begin{center}
S.~J.~LaRoque\altaffilmark{1},
J.~E.~Carlstrom\altaffilmark{1},
E.~D.~Reese\altaffilmark{1},
G.~P.~Holder\altaffilmark{1}, 
W.~L.~Holzapfel\altaffilmark{2}, 
M.~Joy\altaffilmark{3}, 
and L.~Grego\altaffilmark{4}
\end{center}
}
\altaffiltext{1}{Department of Astronomy and Astrophysics, University
of Chicago, Chicago, IL 60637}
\altaffiltext{2}{Department of Physics, University of California,
Berkeley, CA 94720}
\altaffiltext{3}{Department of Space Science, SD50, NASA Marshall Space
Flight Center, Huntsville, AL 35812}
\altaffiltext{4}{Harvard-Smithsonian Center for Astrophysics, 60
Garden Street, Cambridge, MA 02138}

\begin{abstract}
We present an interferometric measurement of the Sunyaev-Zel'dovich
effect (SZE) at 1~cm for the galaxy cluster Abell~2163.  We combine
this data point with previous measurements at 1.1, 1.4, and 2.1 mm
from the SuZIE experiment to construct the most complete SZE spectrum
to date. The intensity in four wavelength bands is fit to determine
the Compton $y$-parameter (\yc) and the peculiar velocity (\vp) for
this cluster.  Our results are $y_{\mbox{\tiny
0}}=3.56^{+0.41}_{-0.41}$$^{+0.27}_{-0.19}\times 10^{-4}$ and
$v_{\mbox{\tiny{\it p}}}=410^{+1030}_{-850}$$^{+460}_{-440} \mbox{ km
s}^{-1}$ where we list statistical and systematic uncertainties,
respectively, at 68\% confidence.  These results include corrections
for contamination by Galactic dust emission.  We find less
contamination by dust emission than previously reported.  The dust
emission is distributed over much larger angular scales than the
cluster signal and contributes little to the measured signal when the
details of the SZE observing strategy are taken into account.

\end{abstract}


\section{INTRODUCTION}
The thermal Sunyaev-Zel'dovich effect (SZE) is a spectral distortion
in the Cosmic Microwave Background (CMB) which results from the
inverse Compton scattering of low-energy CMB photons by thermally
distributed hot electrons in massive clusters of galaxies.  The
thermal SZE manifests itself as an increment in the Wien part of the
CMB spectrum and a decrement in the Rayleigh-Jeans part of the
spectrum \citep{sunyaev72}, with a crossover wavelength of
$\sim1.38$~mm.  There is also a kinetic SZE, caused by the bulk radial
motion of the cluster with respect to the CMB rest frame.  This effect
shifts the CMB Planck spectrum to a slightly lower (higher)
temperature for positive (negative) peculiar velocities where positive
is defined as motion away from the observer.

In this paper, we present a recent 1~cm interferometric detection of
the SZE in the cluster Abell~2163, obtained with the BIMA and OVRO
arrays.  We combine this measurement with a previous result from the
Sunyaev-Zel'dovich Infrared Experiment (SuZIE), a six element
bolometer array (\citealt{holzapfel97b}, hereafter H1997a) on the 10.4
m Caltech Submillimeter Observatory.  The SZE in Abell 2163 was also
observed by Diabolo, a two element bolometer array mounted on the
IRAM 30 m telescope \citep{desert98}, and their result is included in
Table~\ref{tab:one}.  The measurement has considerable uncertainty and
carries very little weight in the spectral fit relative to the SuZIE
data.  Due to the complication of estimating the effects of
systematics on the Diabolo measurement, we have chosen to exclude it
from our spectral fit.  Thus, combining the four measurements from
SuZIE and OVRO/BIMA, we construct the SZE spectrum and use it to
determine the central Compton $y$-parameter and constrain the radial
peculiar velocity of the cluster.  We also address the concern that
Galactic dust along the line of sight to the cluster may be a
significant contaminant in the SuZIE bands \citep{lamarre98}.  Using
data from the Infrared Space Observatory ({\it ISO}) and the Infrared
Astronomical Satellite ({\it IRAS}) we determine the level of dust
contamination in the SuZIE measurements.

In \S2, we describe our observations, data reduction, and dust
analysis.  In \S3 we describe the theoretical model for the SZE
spectrum and present the results of fitting our data to this model; we
also consider several systematic uncertainties affecting the fit.  All
uncertainties are quoted at the 68\% confidence level unless otherwise
specified.

\section{OBSERVATIONS AND DATA REDUCTION}
\label{sec:data}
The 1 cm observations were carried out at the Berkeley Illinois
Maryland Association (BIMA) millimeter array and the Owens Valley
Radio Observatory (OVRO) millimeter array during the summers of 1996,
1997, and 1998.  The telescopes were outfitted with
centimeter-wavelength receivers specifically designed to detect the
SZE in distant galaxy clusters \citep{carlstrom96}.  The receivers use
cooled high electron mobility transistor (HEMT) amplifiers
\citep{pospieszalski95} sensitive to radiation between 0.83 and 1.15~cm
with typical receiver temperatures in the range 11--20~K.
Typical system
temperatures (scaled to above the atmosphere) for the Abell~2163
observations range between 45 and 55~K.

\begin{figure}[tbh]
\epsfysize= 3.0 in
\centerline{\epsfbox{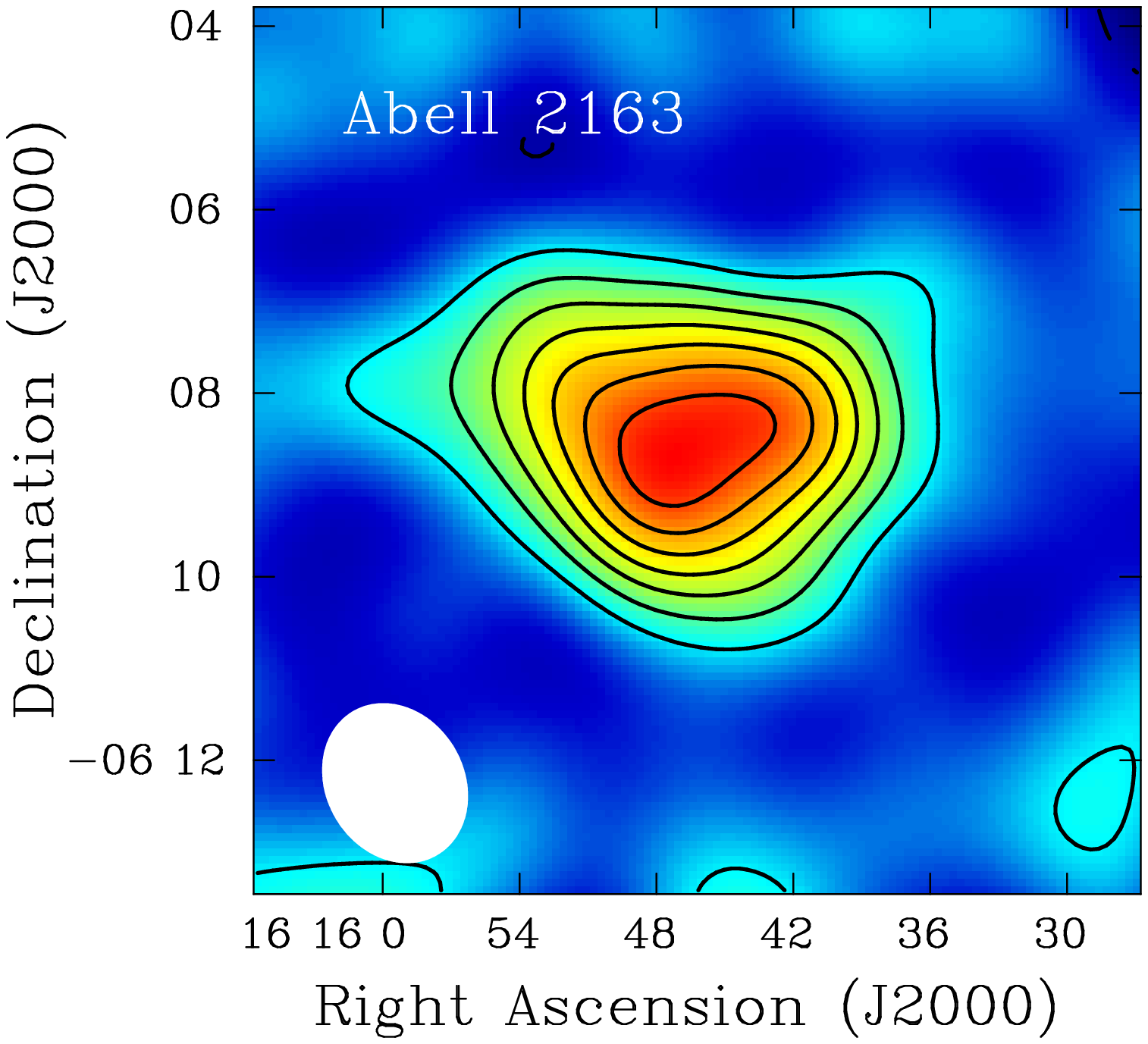}}
\caption{Deconvolved 1 cm SZE image for Abell~2163,
constructed from BIMA data.  The image was made by first applying a
Gaussian taper to the $u$-$v$ data with a half power radius of 1.5
k$\lambda$; the resulting beam FWHM ($110\arcsec \times 90\arcsec$ at
position angle $30^{\circ}$) is shown by the white ellipse in the
lower left.  The rms is $300$~$\mu\mbox{Jy~beam}^{-1}$, corresponding
to a Rayleigh Jeans brightness sensitivity of $50$~$\mu$K.  Contours
are integer multiples of $\pm1.5\sigma$ with negative contours shown
as solid lines.}
\label{fig:sz}
\end{figure}

We observed Abell~2163 at OVRO in 1996 and 1998 for a total of 37
hours with six 10.4 m telescopes providing a $4\arcmin.2$ FWHM primary
beam; we used two 1~GHz wide bands centered at 1.0~cm and 1.05~cm.
This cluster was observed at BIMA in 1996, 1997 and 1998 for 11 hours
with nine 6.1 m telescopes providing a $6\arcmin.6$ FWHM primary beam;
we used an 800~MHz bandwidth centered at 1.05~cm.  At both
observatories, we interleaved cluster measurements with observations
of the radio point source 1549+026 ($3.60\pm0.01$~Jy) every 25 minutes
to monitor system gains, and we used Mars for our absolute flux
calibration \citep{grego00}.  We reduced the data using the MIRIAD
software package \citep{sault95} at BIMA and the MMA package
\citep{scoville93} at OVRO.  In all observations, we flagged data from
baselines where one telescope was shadowed by another, data that were
not bracketed in time by observation of a phase calibrator, and data
showing evidence of poor atmospheric coherence or spurious
correlations.  Figure \ref{fig:sz} shows a deconvolved SZE image of
Abell~2163 constructed from BIMA data.  SZE images are produced with
the software package DIFMAP \citep{pearson94}.  In Figure \ref{fig:sz}
we have applied a Gaussian taper with a half power radius of 1.5
k$\lambda$ to the $u\mbox{-}v$ data.  A bright point source has been
removed from the image in Figure~\ref{fig:sz}; see \S\ref{sec:ptsrcs}
for more details.

An interferometer measures the Fourier transform of the sky
brightness, hence we fit a model to the data in the Fourier plane
where the noise characteristics and spatial filtering of the
interferometer are well understood.  The model is constructed in the
image plane, multiplied by the primary beam of the telescope, then
Fourier transformed to the conjugate plane, known as the $u$-$v$
plane, where it is compared with the data.  

We model the cluster gas as a spherical isothermal-$\beta$ model
\citep{cavaliere78a}, where the SZE intensity change is given by
\begin{equation}
\Delta I_{\nu}(\theta)=\Delta I_{\nu\mbox{\tiny 0}} \left(1+\frac{\theta^2}{\theta_c^2}\right)^{(1-3\beta)/2}.
\label{eq:r_c_beta_I}
\end{equation}
Here \dec\ is the central SZE decrement/increment at frequency $\nu$,
$\theta_c$ is the angular core radius of the cluster, and $\beta$ is a
power law index.  We adopt the model shape parameters
$\theta_c=1\arcmin.20\pm0\arcmin.11$ and $\beta=0.616\pm0.031$ from an
analysis of {\it ROSAT} X-ray data (\citealt{holzapfel97a}, hereafter
H1997b).  With these shape parameters we find \thir$=-0.048\pm0.006
\mbox{ MJy sr}^{-1}$ (corresponding to a CMB temperature decrement
$\Delta T_{\mbox{\tiny CMB}}=-1780\pm210$~$\mu$K) where \thir\
is the SZE decrement at 30~GHz.  The statistical uncertainty in the
central decrement is determined by varying \thir\ while keeping
the shape parameters fixed; all other parameters are free to assume
their best-fit values.  The uncertainties in $\theta_c$ and $\beta$
quoted above lead to an additional $0.001\mbox{ MJy sr}^{-1}$
uncertainty on the 1 cm decrement.

The 1~cm decrement and combined statistical uncertainty are listed in
Table 1, which also contains all spectral SZE data for this cluster
available to date.  The SuZIE \dec\ values are from fits to the SuZIE
data using the average of the measured spectral responses of the
individual channels.  The \dec\ value from the Diabolo experiment is
from \cite{desert98}.  As the table indicates, we have made small
corrections in the SuZIE bands to account for dust contamination along
the line of sight to the cluster.  These corrections are described in
more detail in \S\ref{sec:dust}.  

\subsection{Radio Point Sources}
\label{sec:ptsrcs}
Point sources are identified in DIFMAP using a high-resolution map
produced with data from baselines 20 meters and longer.  We find one
point source in the Abell~2163 data offset $-40\arcsec$ in RA and
$16\arcsec$ in declination from the SZE pointing center, a position
consistent with that of a variable radio source with an inverted
spectrum detected by the VLA (M. Birkinshaw 2001, private
communication) at 2~cm and 6~cm.  The point source has been removed
from the image in Figure \ref{fig:sz}.

The flux and position of the detected point source is jointly fit with
the SZE decrement, but we must consider the uncertainty in the
decrement introduced by point sources in the field with flux densities
below our detection threshold.  Such a point source near the cluster
SZE center would cause us to underestimate the magnitude of the
decrement, whereas a point source in a beam sidelobe could lead to an
overestimate.  To determine the resulting uncertainty we have randomly
added point sources to the data as follows.  Progressing in steps of
1~$\mu$Jy, we choose a Poisson deviate of the expected number of point
sources at each flux density.  We add sources to the field accordingly
out to a radius of $5\arcmin$, well past the half power point of the
BIMA beam where faint point sources should have a negligible effect on
the decrement.  After reaching our $3\sigma$ flux density limit, we
refit the SZE model without accounting for the added point sources.

To determine the number of point sources expected at each flux
density, we first determine a point source number count versus flux
density relationship of the form
\begin{equation}
\label{eq:dnds}
dn/dS \propto S^{-\alpha} \mbox{ mJy}^{-1}\mbox{ arcmin}^{-2} 
\end{equation}
from our sample of 41 clusters observed at BIMA, where $n$ is the
number density of point sources and $S$ is their flux density.  We
find that the distribution of sources beyond $10\arcsec$ from cluster
centers agrees well with the expected distribution from a larger
sample of point sources at 31~GHz (C.~Pryke 2002, private
communication).  Inside $10\arcsec$, however, we detect nearly 30
times more point sources than expected due to the strong contribution
of cluster cD galaxies.  We therefore exclude the inner $10\arcsec$ of
the map when selecting point sources to construct $dn/dS$ and address
the issue of central point sources in more detail below.  We choose
$240\arcsec$ as an outer limit in selecting point sources as it is a
good trade-off between encompassing a large number of detected sources
and allowing a low beam-corrected flux density threshold.  We use only
point sources with beam-corrected flux densities greater than 2~mJy,
since a 2~mJy point source would have a beam-attenuated flux density
of 0.75~mJy at $240\arcsec$ and 0.75~mJy is a good fiducial $3\sigma$
threshold for all 41 maps.  Lastly, we assume the power law index
determined from the Pryke sample, $\alpha=2.4$, in
Equation~(\ref{eq:dnds}).  Solving for the normalization, we find
\begin{equation}
dn/dS=2.0\times 10^{-2} \left(S \over {\rm mJy}\right)^{-2.4}
\mbox{mJy}^{-1} \mbox{ arcmin}^{-2}.
\end{equation}
We then randomly distribute point sources according to this empirical
relationship, starting from a flux density of 10~$\mu$Jy (tests of the
Monte-Carlo showed that starting at a lower flux density has a
negligible effect on the decrement) up to a flux density of 300
$\mu$Jy, corresponding to the $3\sigma$ limit of the more sensitive
OVRO Abell~2163 data.  The resulting distribution of best-fit
decrements is nearly Gaussian and centered at the original decrement
with a standard deviation of 0.002~MJy~sr$^{-1}$.  We adopt this as
the statistical uncertainty due to randomly distributed undetected
point sources, and it has been combined in quadrature with the other
statistical uncertainties in Table~\ref{tab:one}.

We attempt to estimate the uncertainty from undetected point sources a
second way, using information on the distribution of point sources in
the field from observations at longer wavelengths.  Sources with flux
densities greater than $\sim 2$~mJy at 21~cm appear in the NVSS
catalog \citep{condon98}.  The catalog contains four point sources
within $240\arcsec$ of Abell~2163 that we do not detect in our 1~cm
observations.  Searching the 100~$\mu$Jy rms OVRO map at the positions
of these sources, we find no evidence of emission from three of the
four; their positions correspond to signals at the $0.2\sigma$,
$-0.4\sigma$, and $-1\sigma$ levels.  The fourth, however, displaced
$30\arcsec$ in RA and $-127\arcsec$ in declination from the map
center, corresponds to a $2\sigma$ signal in the OVRO map.  We do not
marginalize over the fluxes of all four sources because of the loss of
degrees of freedom in the fit, but marginalizing over the $2\sigma$
source we find no change in either the decrement or the uncertainty in
the decrement, which is dominated by the noise in the map.

To account for the fact that on-center point sources are statistically
more common we place a point source model at the estimated position of
Abell~2163's bright central galaxy (RA
$16^{\mbox{h}}15^{\mbox{m}}49^{\mbox{s}}.0$, Dec
$-6^{\circ}8\arcmin42\arcsec.7$ from Digitized Sky Survey) and
marginalize over its flux.  The new best-fit decrement is
$-0.049$~MJy~sr$^{-1}$, an increase in magnitude of just $0.001\mbox{
MJy sr}^{-1}$ from the original decrement.  The new decrement falls
easily within the 68\% uncertainty of the original; we revisit this
result when we discuss systematic uncertainties (\S\ref{sec:sys_unc}).

\subsection{Analysis of Dust Emission}
\label{sec:dust}
\citet{lamarre98} (hereafter L1998) determined that IR cirrus flux is
a non-negligible contaminant in the three SuZIE bands by including
submillimeter data from the PRONAOS balloon experiment \citep{serra98}
in their fit to the SuZIE and Diabolo data,.  They accounted for dust
by fitting to a combined SZE-spectral model and modified blackbody
model, and found that removing the dust signal changed \dec\
significantly.  Specifically, they found that the 1.1~mm signal
shrinks by $0.051\mbox{ MJy sr}^{-1}$ to $0.236 \mbox{ MJy sr}^{-1}$,
the 1.4~mm signal changes by $0.018\mbox{ MJy sr}^{-1}$ to $-0.124
\mbox{ MJy sr}^{-1}$, and the 2.1~mm band changes by $0.004\mbox{ MJy
sr}^{-1}$ to $-0.387 \mbox{ MJy sr}^{-1}$.  L1998 assumed in their
analysis that the SuZIE scan pattern is identical to the PRONAOS
chopping scheme.  However, PRONAOS performed a $6\arcmin.2$ chop at
constant elevation (see L1998 for details) while SuZIE uses a drift
scan at constant declination (H1997b).  Thus, the experimental filters
are different and will not necessarily detect the same dust signal.

To better estimate the level of dust contamination in the high
frequency SuZIE channels, we have applied the SuZIE observing scheme
to extrapolated dust maps at the SuZIE wavelengths.  The expected dust
contamination is listed in Table~\ref{tab:two} and the dust-corrected
SuZIE measurements are given in Table~\ref{tab:one}.  Here we describe
the procedure.

For clarity, we summarize the map construction procedures as applied
to the 1.1~mm band; the 1.4~mm and 2.1~mm procedures are identical.
The 1.1~mm dust map is created by extrapolating from three existing
dust maps centered on Abell~2163; {\it ISO} maps at 90~$\mu$m and
180~$\mu$m, and an {\it IRAS} map at 100~$\mu$m.  We extract a
$15\arcmin \times 1\arcmin.75$ central region from each dust map,
consistent with the area of the SuZIE scans.  We then smooth each map
to a resolution consistent with the SuZIE beam size ($1\arcmin.75$),
taking into account the intrinsic resolution of the IRAS and ISO
experiments.  We use a fiducial pixel to pixel uncertainty of $0.2
\mbox{ MJy sr}^{-1}$ (cf.\ \citealt{beichman88}) for the {\it IRAS}
map; for the {\it ISO} maps, we adopt the values estimated in L1998 of
$0.14 \mbox{ MJy sr}^{-1}$ and $0.18 \mbox{ MJy sr}^{-1}$ at 90 and
180~$\mu$m, respectively.  For each {\it ISO} map, we then bin the
$0\arcmin.25$ pixels into $0\arcmin.75$ pixels, consistent with
binning performed in the SuZIE analysis.  For each new $0\arcmin.75$
pixel, we take the average of the original three $0\arcmin.25$ pixels
as the intensity.  The {\it IRAS} pixels are roughly twice the size of
the binned {\it ISO} pixels, so we linearly interpolate to estimate
the intensity of interleaved pixels, which we artificially add to the
{\it IRAS} map.  We then fit a modified blackbody to these dust maps
at each pixel position; since we have three dust maps, each fit is to
three data points, or one ``pixel triplet''.  This allows us to
extrapolate the intensities out to 1.1~mm.  We fit a modified
blackbody of the form
\begin{equation}
F_d \propto \frac{\nu^{3+\gamma}}{e^{h\nu/kT_d}-1},
\label{eq:mod_bb}
\end{equation}
where $\gamma$ is the spectral index, $T_d$ is the dust temperature,
and we normalize the model with the dust flux at 180~$\mu$m,
$F_d(180)$.  Following L1998, we fix the spectral index at $\gamma=2$.
Upon fitting each dust pixel triplet to this model, we find $F_d(180)$
in the range $25.1$ to $26.1$~MJy sr$^{-1}$ and $T_d$ in the range
21.3 to 21.9~K over all triplets.  We then use each triplet's best fit
$F_d(180)$ and $T_d$ with Gaussian error propagation to extrapolate
the intensity at that position and create our 1.1~mm dust map.

After repeating this process at 1.4 and 2.1~mm, we perform the
simulated scan following the detailed discussion of SuZIE's
instrumental specifications and observing scheme in H1997b.  In our
simulation we scan across the $0\arcmin.75$ pixels of the dust maps in
the direction of increasing right ascension (RA).  For each pixel
along the scan, we record the difference between its intensity and
that of the pixel advanced $4\arcmin.6$ in RA; pixel-to-pixel
uncertainties are combined in quadrature.  This procedure is
equivalent to SuZIE allowing the sky to drift overhead with the
earth's rotation and recording the difference in signal between two
bolometer array elements separated by $4\arcmin.6$ on the sky.  We do
not treat SuZIE's $2\arcmin.3$ ``triple beam chop'' here as it has
little sensitivity to linear changes in brightness across the sky and
therefore is less sensitive to dust; the results presented here serve
as an upper limit to the dust contribution (see H1997b for more
details).  Our simulation thus creates ``differenced'' maps of the
dust brightness.

We perform one-dimensional $\chi^2$ fits in the parameter \dec\ to the
dust maps, fitting to a SuZIE-differenced $\beta$ model with the
pixel-to-pixel dust noise as our uncertainty.  We use a Monte Carlo
procedure and successively fit to $10^6$ different realizations of the
dust; in a single realization, each pixel has added to it some
Gaussian-distributed amount of the pixel-to-pixel noise.  The results
of the Monte Carlo are highly Gaussian as expected; we thus take the
most frequently returned \dec\ as our best fit and those that enclose
$68.3\%$ of the returned \dec's as our $1\sigma$ uncertainties.  The
results are shown in Table~\ref{tab:two} and reflected in the last
column of Table~\ref{tab:one}.  We find a dust contamination level
roughly seven times smaller than that reported in L1998, and small
compared to the statistical uncertainties of the SZE measurements.
 
\section{SZE SPECTRAL MODEL AND RESULTS}
\label{sec:modres}
We have fit the dust-corrected SZE spectral data in Table
\ref{tab:one} to a model which consists of both the thermal and
kinetic SZE components.  The thermal component, $\Delta I_{\mbox{\tiny
T}}$, can be written as
\begin{equation}
\Delta I_{\mbox{\tiny T}} = I_{\mbox{\tiny 0}} y f(x) (1+\delta_{\mbox{\tiny T}})
\label{eq:I_t}
\end{equation}
where $x=h\nu/kT_{\mbox{\tiny CMB}}$, $T_{\mbox{\tiny CMB}}=2.728$~K
is the CMB temperature \citep{fixsen96}, $\delta_{\mbox{\tiny T}}$ is
a relativistic correction to the thermal effect good to fifth order in
$kT_e/m_e c^2$ \citep{itoh98}, $I_{\mbox{\tiny 0}} \equiv
2(kT_{\mbox{\tiny CMB}})^3/(hc)^2$, and
\begin{equation}
f(x) = \frac{x^{4} e^{x}}{(e^{x} - 1)^{2}}
\left[\frac{x (e^{x} +1)}{e^{x} -1} - 4\right].
\end{equation}
The quantities $h$, $k$, and $c$ are the Planck constant, Boltzmann
constant, and speed of light, respectively.
The quantity $y$ is the Compton $y$-parameter, and is given by
\begin{equation}
y=\int \left(\frac{kT_e}{m_e c^2} \right) n_e \sigma_T dl,
\label{eq:y_c}
\end{equation}
where $T_e$ is the electron temperature of the
intracluster medium (ICM),
$m_e$ is the electron mass, $n_e$ is the electron density, and $\sigma_T$
is the Thomson scattering cross section.  The Compton $y$-parameter is
proportional to the pressure integrated along the line of sight ($dl$)
through the cluster; we parameterize it in terms of a central
value, which we denote \yc.

The kinetic component, $\Delta I_{\mbox{\tiny K}}$, depends on both
$y$ and the radial peculiar velocity of the cluster, \vp, through the
equation
\begin{equation}
\Delta I_{\mbox{\tiny K}} = -(m_ec^2/kT_e) I_{\mbox{\tiny 0}} y
\frac{v_{\mbox{\tiny{\it p}}}}{c} g(x) (1 + \delta_{\mbox{\tiny K}})
\label{eq:I_k}
\end{equation}
where $g(x)=x^4 e^x / (e^x -1)^2$ and $\delta_{\mbox{\tiny K}}$ is a
relativistic correction to the kinetic effect good to third order in
$kT_e/m_e c^2$ \citep{nozawa98}.  Corrections of order \vp$^2/c^2$ to
the thermal and kinetic SZE are neglected since \vp$\ll c$.  The total
intensity is then the sum of the thermal and kinetic components,
\begin{equation}
\Delta I_{\nu}=\Delta I_{\mbox{\tiny T}} + \Delta I_{\mbox{\tiny K}}.
\label{eq:total_I}
\end{equation}
We perform a least squares fit to this model in the parameters \yc\
and \vp\ assuming an ICM temperature $T_e$ of $12.4^{+2.8}_{-1.9}$ keV
from H1997b.

The resulting \vp\ and \yc\ from the least squares fit to the four
data points are $y_{\mbox{\tiny 0}}=3.56\pm0.41\times 10^{-4}$ and
$v_{\mbox{\tiny{\it p}}}=410^{+1030}_{-850} \mbox{ km s}^{-1}$ with
$\chi^2=0.8$ for two degrees of freedom.  We compare this with the
results of our fit to the SuZIE data only, $y_{\mbox{\tiny
0}}=3.72\pm0.51\times 10^{-4}$ and $v_{\mbox{\tiny{\it
p}}}=290^{+1020}_{-820} \mbox{ km s}^{-1}$.  The addition of the 1~cm
point improves the \yc\ constraints by 18\%.  The constraints on \vp\
hardly change when the 1~cm point is included since constraints on
peculiar velocity are dominated by the SuZIE data points which
straddle the thermal SZE null.

Figure \ref{fig:SZE-spectrum} shows the SZE spectrum generated from
the fit to all four frequency bands.  There is excellent consistency
amongst all of the measurements.  The data were obtained from
independent experiments, each with its own telescope, detector, and
observing strategy, operating at four different wavelengths.  The
consistency amongst the data attests to the reliability and accuracy
with which the SZE is now being measured.

\subsection{Systematic Uncertainties}
\label{sec:sys_unc}
In this section we address several additional sources of uncertainty
in our spectral fits; the results of this analysis are summarized in Table
\ref{tab:three}.

We first consider the systematic uncertainty associated with the dust
emission calculated in \S\ref{sec:dust}.  When we alter the dust
intensity at the 68\% level, the spectral fit results change by
$^{+0.00}_{-0.01}\times10^{-4}$ for \yc\ and $^{+0}_{-30}\mbox{ km
s}^{-1}$ for \vp.  The dust uncertainties are sufficiently small as to
have very little impact on the spectral fit results.  We find a more
significant change when we repeat the dust analysis with a smaller
spectral index, $\gamma$.  \citet{finkbeiner99} recently suggested
based on fits to FIRAS data, that while $\gamma=2$ is appropriate for
wavelengths $\lesssim 600$~$\mu$m, the spectral index at longer
wavelengths is closer to 1.5.  Setting $\gamma=1.5$ for the entire
dust spectrum we find that \yc\ increases by $0.02\times 10^{-4}$ and
\vp\ decreases by $70\mbox{ km s}^{-1}$.  We combine these in
quadrature with the changes that resulted from scaling the dust at the
68\% level and adopt the total as the systematic uncertainty due to
dust contamination.

Temperature fluctuations in the CMB are a potential source of
confusion for SZE measurements.  Measurements of the kinetic SZE are
particularly susceptible since the spectral dependence of the kinetic
SZE and CMB fluctuations are identical, absent relativistic
corrections to the SZE at the few percent level.  Recent BIMA
observations at 28.5 GHz have provided good constraints on CMB
anisotropy at the angular scales of our measurement
\citep{dawson01,holzapfel00a}; they find $\Delta I_{\nu} < 4.8 \times
10^{-4} \mbox{ MJy sr}^{-1}$ at 95\% confidence, which is just 1\% of
\thir.  This uncertainty is negligible given the precision of our
measurement.  The level of CMB contamination may be larger in the
SuZIE bands, however, particularly near the null in the thermal SZE.

To test the level of CMB contamination in the SuZIE bands we first use
the software package CMBfast \citep{zaldarriaga00} to generate a CMB
power spectrum up to a multipole $\ell=2500$.  We assume a universe
consisting of cold dark matter with a cosmological constant
($\Lambda$CDM) and use conventional parameter values
$(\Omega_b,\Omega_{cdm},\Omega_{\Lambda},\tau_c,n_s,h)=
(0.04,0.26,0.70,0,0.95,0.72)$ where $\Omega_b$ is the fractional
baryon density of the universe, $\Omega_{cdm}$ is the fractional cold
dark matter density, $\Omega_{\Lambda}$ is the fractional density of
vacuum energy, $\tau_s$ is the optical depth to reionization, $n_s$ is
the spectral index of the power spectrum, and
$H_0=100h$~km~s$^{-1}$~Mpc$^{-1}$ is the Hubble constant.  The values
chosen are good estimates given the range of values currently
constrained by CMB anisotropy measurements
\citep{pryke01,netterfield01,stompor01}; $h$ comes from the Hubble
Space Telescope Hubble Constant key project \citep{freedman01}.  To
account for secondary anisotropies we conservatively assume a flat
power spectrum for $\ell>2500$, corresponding to arcminute angular
scales.  The flat band power we use corresponds to $\Delta
T_{\mbox{\tiny CMB}}/T_{\mbox{\tiny CMB}}\sim 5\times10^{-6}~$, which
is consistent with expectations of secondary anisotropies \citep{hu01}
and the BIMA limit of \citet{dawson01}.  We then use this power
spectrum to produce mock CMB fields at SuZIE's resolution.  After
simulating the SuZIE scan on these maps, we fit them to an isothermal
$\beta$-model and use a Monte Carlo procedure similar to that used
with the dust to determine the contribution of CMB to the total signal
in each SuZIE band.  Upon fitting to $10^4$ different maps for each
band, we find an rms of $\pm0.020\mbox{ MJy sr}^{-1}$ at 1.1~mm,
$\pm0.022\mbox{ MJy sr}^{-1}$ at 1.4~mm, and $\pm0.017\mbox{ MJy
sr}^{-1}$ at 2.1~mm.  Further tests using only primary anisotropy
$(\ell \lesssim 2500)$ or only secondary anisotropy $(\ell \gtrsim
2500)$ indicate that primary anisotropy is responsible for roughly
80\% of the total CMB contribution in all three bands.  Including the
rms total anisotropy contributions in the SuZIE signals listed in
Table \ref{tab:one}, we redo the spectral fit and find changes of
$\pm0.05$ in \yc\ and $^{+420}_{-400}\mbox{ km s}^{-1}$ in \vp.  The
peculiar velocity suffers more from confusion with CMB anisotropy than
does the Compton $y$-parameter, as anticipated.  We adopt these
results as the systematic uncertainties due to CMB contamination.

We also consider confusion with extended emission from the radio halo
surrounding Abell~2163, which is the brightest such halo yet
discovered \citep{feretti01}.  To test its effect on the 1~cm
decrement, we combine the OVRO data with the 1.4~cm NVSS map of the
cluster region, which shows a large region of extended emission
surrounding the cluster.  We first scale the NVSS map to account for
the unknown spectrum of the radio halo; we make this scale factor a
free parameter and redo the fit described in \S\ref{sec:data}.  The
decrement is increased by $-0.003 \mbox{ MJy sr}^{-1}$ when we include
this extra parameter, and the scaling is consistent with a spectral
index $\alpha\simeq1.2$ where the intensity has a $\nu^{-\alpha}$
dependence.  We note, however, that our fit is consistent with
$\alpha=\infty$ (i.e., no radio halo) within 68\% confidence.
Furthermore, the spectral index may not be constant across the halo,
so the best-fit $\alpha$ must be taken as a crude estimate.  H1997b
also address radio halo confusion, and estimate a 2\% effect at
2.1~mm.  This assumes a spectral index of $\sim 1.5$, however; using
our best-fit value $\alpha=1.2$, the effect increases to $\sim 5\%$.
When we alter the 1~cm and 2.1~mm points accordingly and redo the
spectral fit, we find that \yc\ increases by $0.17\times10^{-4}$ to
$3.73\times10^{-4}$, and \vp\ increases by $50\mbox{ km s}^{-1}$ to
$460 \mbox{ km s}^{-1}$.  We adopt these as estimates of the
systematic uncertainties in \yc\ and \vp\ due to radio halo confusion.

We use a similar treatment to determine the systematic uncertainty due
to point source contamination.  In \S\ref{sec:data} we accounted for
the possibility of an undetected on-center point source in the field
at 1~cm, and found that it could increase the magnitude of the
measured decrement by $0.001 \mbox{ MJy sr}^{-1}$.  Such a faint
source would not affect the SuZIE data unless it had a strongly
inverted spectrum; we assume that it does not.  The brighter
off-center point source detected at 1~cm could affect the SuZIE data
if its spectrum were inverted at millimeter wavelengths.  Although its
flux rises from 1 to 3~mJy between 6~cm and 2~cm (M. Birkinshaw 2001,
private communication), we found at the time of our observations that
the flux of the source at 1~cm varied between 1.1 and 1.4~mJy.  This
corresponds to a spectral index $\alpha$ between 1.1 and 1.5.
Extrapolating with such a spectral index to 2.1~mm, we find a flux of
less than 0.2~mJy which would have a negligible effect on the
decrement.  However, because of its variability and unusual spectrum,
we conservatively assume a flat spectral index ($\alpha=0$) for
$\lambda < 1$~cm.  Changing the 1~cm decrement to account for a faint
central point source and altering the SuZIE 2.1~mm decrement to
account for a 1.5~mJy point source, we find an increase of
$0.07\times10^{-4}$ in \yc\ and $50 \mbox{ km s}^{-1}$ in \vp.  We
include these in Table \ref{tab:three} as our systematic uncertainty
due to point source contamination.

We test the importance of our assumed 12.4 keV electron temperature to
\yc\ and \vp\ by performing the spectral fit over a large range in
$T_e$.  An increase in $T_e$ trades off with a decrease in the cluster
optical depth $\tau=\int n_e \sigma_T dl$, as can be seen in equation
(\ref{eq:y_c}); thus so long as $\tau$ remains a free parameter,
Compton $y$ is only weakly sensitive to temperature changes.  In the
$1\sigma$ range of $T_e$ ($10.5 \mbox{ keV}\leq T_{\mbox{\scriptsize
e}} \leq 15.2 \mbox{ keV}$), \yc\ varies by
$^{+0.09}_{-0.07}\times10^{-4}$.  The peculiar velocity also has a
weak temperature dependence; it rises with temperature at low
$T_{\mbox{\scriptsize e}}$ ($\lesssim 5 \mbox{ keV}$) as $\tau$
decreases (cf.\ equation \ref{eq:I_k}), then falls at high $T_e$
($\gtrsim 15 \mbox{ keV}$) due to relativistic corrections.  The
$1\sigma$ range of $T_e$ corresponds to a broad plateau in the
$T_e$-\vp\ curve, and we find \vp\ varying by only $+20\mbox{ km
s}^{-1}$ and $-40\mbox{ km s}^{-1}$.

We have assumed an isothermal temperature distribution for the cluster
gas, consistent with the H1997a and b treatments of Abell~2163.
Although the assumed cluster temperature has little impact on \yc\ and
\vp, uncertainties in the thermal structure of the cluster may have a
larger effect.  In a recent study of Abell~2163 using the {\it
BeppoSAX} X-ray satellite, \citet{irwin00} find the cluster
temperature profile consistent with isothermality at the $2\sigma$
confidence level out to a radius of $9\arcmin$; this is consistent
with the result of \citet{white00}, who finds a temperature profile
consistent with isothermality at the $1\sigma$ confidence level out to
$10\arcmin$.  Chandra or XMM data are necessary to place stronger
constraints on the thermal profile of Abell~2163, and we postpone
quantifying a systematic uncertainty until such data become available.

Finally, we consider the effect of the assumed $\pm8\%$ absolute
calibration uncertainty in each SuZIE band and the $\pm3\%$ absolute
calibration uncertainty in our interferometric 1~cm measurements.
These combine to change \yc\ by as much as
$\pm0.17\times10^{-4}$ and \vp\ by up to $\pm170\mbox{ km
s}^{-1}$.

\section{CONCLUSIONS}
We have used interferometric data obtained at the OVRO and BIMA
observatories to determine the SZE intensity decrement in the cluster
Abell~2163 at a wavelength of 1~cm.  We find \thir$=-0.048\pm0.006$
MJy sr$^{-1}$ by modeling the ICM as an isothermal spherical
$\beta$-model.  We have thus expanded the spectral coverage of this
cluster to include the Rayleigh-Jeans part of the CMB blackbody
spectrum and we present the most complete SZE spectrum to date.  Upon
fitting the OVRO/BIMA and SuZIE data for Abell~2163 to an SZE spectral
model, we find $y_{\mbox{\tiny
0}}=3.56^{+0.41}_{-0.41}$$^{+0.27}_{-0.19}\times 10^{-4}$ and
$v_{\mbox{\tiny{\it p}}}=410^{+1030}_{-850}$$^{+460}_{-440} \mbox{ km
s}^{-1}$ where we list statistical uncertainty followed by systematic
uncertainty at 68\% confidence.  The systematic uncertainties in Table
\ref{tab:three} have been combined in quadrature to determine the
totals reported above.  We have shown that adding the 1~cm measurement
improves constraints on the Compton $y$-parameter, as expected, but
has less effect on the peculiar velocity which is best constrained by
measurements near the 218~GHz thermal null.

The spectral fit includes corrections for dust emission in the SuZIE
bands; we find that the contamination level depends strongly on the
observing scheme of the instrument due to the distribution of dust on
the sky.  The dust contamination level for a cluster will obviously
depend on its location on the sky; the line of sight to Abell~2163
passes near the Galactic plane, so it is more susceptible to dust
contamination than clusters at higher galactic latitudes.

\begin{figure}[tbh]
\epsfysize= 3.0 in
\centerline{\epsfbox{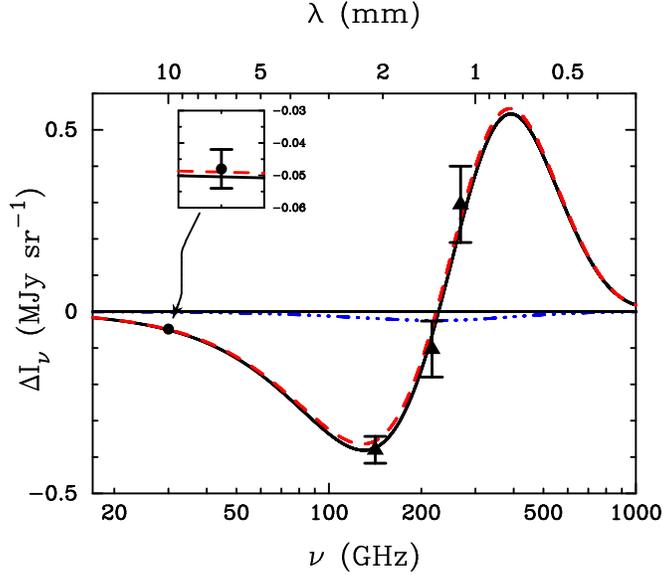}}
\caption{SZE spectrum of Abell~2163 (points) and best fit
model (lines).  The dashed line is the thermal spectrum, dot-dashed
the kinetic spectrum, and solid the sum of the two; SuZIE data points
appear as triangles and OVRO \& BIMA as a circle.  The insert shows
the 1 cm point with its error bar.}
\label{fig:SZE-spectrum}
\end{figure}

Finally, we emphasize that the Abell~2163 spectrum shown in Figure
\ref{fig:SZE-spectrum} consists of data from two very different
experiments; an array of six bolometers performing drift scans (SuZIE)
and two interferometers (OVRO and BIMA).  The remarkable consistency
that we find amongst the two sets of measurements is further
evidence that SZE observations are now a reliable probe of structure
in the universe.

\acknowledgements This work is supported by NASA LTSA grant
NAG5--7986.  Many thanks to Erik Leitch for the use of his CMB
field-generating software and for many helpful discussions.  Thanks
also to Amber Miller for useful discussions and much help with point
sources.  Clem Pryke provided invaluable information on the flux
distribution of point sources detected by the DASI experiment.  We
thank the staff of the OVRO and BIMA observatories for their support
and hard work, in particular J.\ R.\ Forster, J.\ Lugten, S.\ Padin,
R.\ Plambeck, S.\ Scott, and D.\ Woody.  Radio astronomy at the BIMA
millimeter array is supported by NSF grant AST 96--13998, and the OVRO
millimeter array is supported by NSF grant AST 96--13717.  E.\ R.\
acknowledges support from NASA GSRP Fellowship NGT5--50173, and J.\
C.\ acknowledges support from a NSF--YI grant and the David and Lucile
Packard Foundation.  The Digitized Sky Survey was produced at the
Space Telescope Science Institute under U.S. Government grant NAG
W-2166. The images of these surveys are based on photographic data
obtained using the Oschin Schmidt Telescope on Palomar Mountain and
the UK Schmidt Telescope. The plates were processed into the present
compressed digital form with the permission of these institutions.  We
thank S. Hansen, S. Pastor, and D. Semikoz for politely pointing out
an error in our initial spectral fit.  Finally, we thank M. Birkinshaw
and M. Giard for the useful information they have provided.
     
\newpage


\clearpage

\newpage

\begin{deluxetable}{lclc}
\tablewidth{0pt}
\tablenum{1}
\tablecaption{Measurements of SZE in Abell~2163  \label{tab:one}}
\tablehead{
  \colhead{$\lambda$} &
  \colhead{Instrument} &
  \colhead{Measured \dec} &
  \colhead{Dust-corrected \dec}\\
  \colhead{(mm)} &
  \colhead{} &
  \colhead{(MJy sr$^{-1}$)} &
  \colhead{(MJy sr$^{-1}$)}
}
\tablecolumns{4}
\startdata
10.0 & OVRO \& BIMA  & $-0.048\pm0.006$ &\ldots\\
2.1 & Diabolo &  $-0.545\pm0.22$ &\ldots\\
2.1 & SuZIE &  $-0.381\pm0.037$ & $-0.380\pm0.037$\\
1.4 & SuZIE &  $-0.106\pm0.077$ & $-0.103\pm0.077$\\
1.1 & SuZIE &  $\mbox{\phm{$-$}}0.287\pm0.105$ &
$\mbox{\phm{$-$}}0.295\pm0.105$
\enddata
\end{deluxetable}

\begin{deluxetable}{cc}
\tablewidth{0pt}
\tablenum{2}
\tablecaption{Strength of Dust Signals \label{tab:two}}
\tablehead{
  \colhead{$Map$} &
  \colhead{\dec}\\
	\colhead{($\mu$m)} & \colhead{($\mbox{MJy sr}^{-1}$)}	
}
\tablecolumns{2}
\startdata
1100 & $-7.6^{+1.2}_{-1.4}\times10^{-3}$ \\
1400 & $-3.5^{+0.5}_{-0.6}\times10^{-3}$ \\	
2100 & $-0.7^{+0.1}_{-0.1}\times10^{-3}$	
\enddata
\end{deluxetable}

\begin{deluxetable}{lrr}
\tablewidth{0pt}
\tablenum{3}
\tablecaption{Systematic Uncertainties in \yc\ and \vp \label{tab:three}}
\tablehead{
  \colhead{Source} &
  \colhead{Effect on \yc} &
  \colhead{Effect on \vp}\\
	\colhead{} & \colhead{($\times10^{-4}$)} & 
	\colhead{($\mbox{km s}^{-1}$)}
}
\tablecolumns{3}
\startdata
Dust Contamination & $(+0.02,-0.01)$ & $(+30,-70)$ \\
CMB Anisotropy & $(+0.05,-0.05)$ & $(+420,-400)$ \\
Radio Halo & $(+0.17,-0)$ & $(+50,0)$ \\
Point Sources & $(+0.07,-0)$ & $(+50,-0)$ \\
X-ray Temperature & $(+0.09,-0.07)$ & $(+20,-40)$ \\
Absolute Calibration & $(+0.17,-0.17)$ & $(+170,-170)$ \\
 & &  \\
Total\tablenotemark{a} & $(+0.27,-0.19)$ & $(+460,-440)$ \\  
\enddata
\tablenotetext{a}{combined in quadrature.}
\end{deluxetable}

\end{document}